\documentclass[aps,prd,twocolumn,showpacs,superscriptaddress]{revtex4}

\usepackage{dcolumn}
\usepackage{bm}
\usepackage{graphicx}

\begin{document}

\title{Observing Nucleon Decay in Lead Perchlorate}

\author{R.~N.~Boyd} 
\affiliation{Department of Physics, Ohio State University, Columbus, 
OH 43210}
\affiliation{Department of Astronomy, Ohio State University, Columbus, 
OH 43210} 
\author{T.~Rauscher}
\affiliation{Department of Physics, University of Basel, CH-4056 Basel, 
Switzerland}
\author{S.~D.~Reitzner}
\affiliation{Department of Physics, Ohio State University, Columbus, 
OH 43210}
\author{P.~Vogel}
\affiliation{Department of Physics, California Institute of Technology,
Pasadena, CA 91125}

\date{\today}

\begin{abstract}
Lead perchlorate, part of the OMNIS supernova neutrino detector, contains
two nuclei, $^{208}$Pb and $^{35}$Cl, that might be used to study nucleon
decay. Both would produce signatures that will make them especially useful
for studying less-well-studied neutron decay modes, \textit{e.g.}, those in
which only neutrinos are emitted. 
\end{abstract}


\maketitle

\section{Introduction}

Studies of nucleon decay (see, \textit{e.g.}, \cite{pati73,georgi74, 
babu00,babu01,applequist01}) are among the most important in physics, in
that they provide direct tests of fundamental theories of particle physics.
The different particle theories make rather different
predictions as to what the nucleon decay half-lives might be, and even
which decay modes would be expected to dominate. Although a number of
searches for nucleon decay have been performed in the large detectors that
exist, no convincing evidence has yet been presented for its existence. The
best limits from these studies have involved decay modes in which protons
decay to relativistic charged leptons (see, \textit{e.g.}, \cite{pdg}),
either as individual protons or as protons in nuclei, as the resulting
Cherenkov radiation would produce definitive signatures, typically of order
10$^{33}$ y. However, nucleon decays in composite nuclei \cite{totsuka86,
ejiri93,suzuki93,kamyshkov02} might 
produce clear evidence for the existence of such effects that could not
be obtained from decay of isolated protons. Furthermore, these might allow
extension of the existing limits of the decay branches in some instances,
even with relatively small detectors.

In this study we focus on the n $\rightarrow$ $\nu \ + \ \bar{\nu} \ + \
\nu$ decay, which has a less well-known half-life because its decay can
often be masked by backgrounds, and would in any event be difficult to
detect via the means used to observe nucleon decay into relativistic
leptons. It is not necessarily expected to be the dominant mode of decay,
but it is the least-well-determined mode, so is the primary limitation to
the mode-independent half-life. Moreover, there are non-standard models in
which unusual decay modes may turn out to be the dominant ones (see, {\it
e.g.}, \cite{pati83}). In some models the three neutrino decay mode can
emerge naturally as the dominant one \cite{mohapatra02}.

Decays of the three neutrino mode were searched for a decade ago,
\cite{berger91}, then their limit was improved to its current experimental
value of 5x10$^{26}$ y \cite{pdg}. Recent suggestions for studying this
decay mode have included signatures that would result from decay of a
neutron in the O in H$_2$O \cite{suzuki93} of Super-Kamiokande and in the C
in the CH$_2$ \cite{kamyshkov02} of KamLAND. These suggestions both involve
signals that would be generated by rather weak branches resulting from
nucleon decay.

\section{Signals from Lead Perchlorate}

In this paper we study two nuclides, $^{208}$Pb and $^{35}$Cl which, we
show, would produce special responses to neutron decay. We present the
signatures that might result in the n $\rightarrow$ $\nu \ \bar{\nu} \ \nu$
decay. We find that $^{208}$Pb would have good sensitivity to this mode, but
$^{35}$Cl, while not producing as strong a limit, would exhibit an unusually
definitive signature. Both nuclei are part of a planned detector, lead
perchlorate, LPC, Pb[ClO$_4$]$_2$, that will be part of OMNIS, the
Observatory for Multiflavor NeutrInos from Supernovae \cite{boyd02}. LPC is
a colorless liquid that is highly soluble in water. The properties of such a
mixed liquid have been studied extensively \cite{elliott00}; it was found
that the LPC will produce Cherenkov light from relativistic particles, e.g.,
electrons and muons, that might be produced in many nucleon decays. In
addition, each neutron produced by neutron emission from a nucleus within
the detector will be captured on the $^{35}$Cl within tens of $\mu$s,
producing 8.6 MeV of $\gamma$-rays. An energetic $\gamma$-ray
will also produce Cherenkov light, with a pattern that is indistinguishable
from that from the relativistic leptons in LPC.

In general we can write the following expression for the neutron decay
lifetime, $\tau_n$, as 
\begin{equation}
\tau_n/Br > N_n \epsilon_n R_{det} \epsilon_1 \epsilon_2^n,
\end{equation}
where N$_n$ is the number of neutrons in, \textit{e.g.}, $^{208}$Pb or 
$^{35}$Cl, $\epsilon_n$ is the fraction of those that can decay into
detectable signatures, R$_{det}$ is the observed event rate, and
$\epsilon_1$ and $\epsilon_2$ are the detection efficiencies for the two (or
more) signatures of the decay. The factor $\epsilon_2^n$ accounts for the
possibility of multiple neutron emission from decay of a neutron in
$^{208}$Pb, each with detection efficiency $\epsilon_2$. The factor Br is
the branching ratio for the decays that go to the specific decay mode being
studied. In $^{208}$Pb decay, this could, 
\textit{e.g.}, refer to the branching ratio for producing one neutron and a
subsequent $\gamma$-ray with sufficient energy to be observed; 
we have assumed E$_{ex}$ $<$ 3 MeV in the daughter nucleus for this
criterion to be satisfied, as virtually every level with excitation energy
above that will produce at least one 3 MeV $\gamma$-ray
\cite{nucdatasheets}. In $^{35}$Cl decay, Br refers to a decay to states
that will produce a $\gamma$-ray with at least 3 MeV of energy together with
decay to the $^{34}$Cl ground state. ($^{34}$Cl
has an isomeric state at 0.146 MeV, to which roughly half of the highly 
excited states will ultimately decay. However, it has a much longer
half-life than the 1.5 s half-life of the ground state, so will
be assumed not to be useful for the present discussion.)

In order to estimate the probability of observing decays from $^{208}$Pb and
$^{35}$Cl we will have to consider the probability of population of
sufficiently highly excited states in the daughter nuclei: $^{207}$Pb and
$^{34}$Cl. Generally the more
deeply bound the nucleon that decays the larger will be the excitation
energy in the daughter nuclide. The nuclide that decays will be a nuclide in
the parent nucleus, so the resulting ``state" of the initial nucleus,
$^{208}$Pb or $^{35}$Cl, minus one neutron will map onto actual states in
the daughter nuclide (assuming the neutron decay leaves the resulting
nucleus intact; a reasonable assumption for the decay mode that produces
three particles that interact only through the weak interaction). However,
consider the energy $E_{decay}$ of the
neutron decay products (e.g. the 3 neutrinos) when the final nucleus A-1 is
excited to the excitation energy $E_{exc}$. Then $E_{decay} = M_n - (S_n +
E_{exc})$, where $S_n$ is the neutron separation energy in the 
parent nucleus A. The energy in parentheses is just the binding energy of
the neutron that decayed. Then, all other things being equal, extremely
large values of $E_{exc}$ will be suppressed by the decrease of the phase
space of the 3 neutrinos. This will tend to favor the excitation energies of
the states in the daughter being similar to the binding energy of the
neutron that decayed. Of course, this argument also assumes that the recoil
kinetic energy of the residual nucleus will be negligible.

\subsection{The n $\rightarrow \ \nu \ \bar{\nu} \ \nu$ decay mode in
$^{208}$Pb.}

As discussed above, the signature of this nucleon decay would be given by
the specific properties of the residual nuclei resulting from the decay. In
this case, $^{208}$Pb would first become $^{207}$Pb. What would happen
next would depend on the excitation energy of the states in $^{207}$Pb that
were populated compared to the one-neutron emission threshold, the
two-neutron emission threshold, etc., in $^{207}$Pb. Subsequent neutron
emissions will be much more rapid than electromagnetic de-excitations, so
subsequent neutron emissions would occur instead of de-excitations of the
daughter nuclei as long as the states populated were above the neutron-decay
threshold. The excitation energy in the post-decay $^{207}$Pb will depend on
the binding energy of the nucleon that decayed. This energy would be
expected to be as much as several tens of MeV in lead, but would be weighted
toward lower values by the preponderance of higher spin nuclear orbits,
hence higher occupation numbers, near the Fermi surface. Thus a reasonable
range to assume for the excitation energy in $^{207}$Pb might be an
asymmetric distribution ranging from essentially zero (if a valence neutron
decayed) to as much as 20 MeV, with an even higher-energy tail of the
distribution extending to several more tens of MeV, but with the highest
energies suppressed. The one-neutron (two-neutron) emission threshold in
$^{207}$Pb is 6.74 MeV (14.83 MeV). Decays of neutrons in the highest-energy
occupied shells would presumably primarily populate states below the
one-neutron-emission threshold in $^{207}$Pb. Although these states would
decay by emitting $\gamma$-rays, $^{207}$Pb is stable. Thus the
$\gamma$-rays would be the only signature of the neutron decays to the
low-lying states, providing a less-than-compelling signature of nucleon
decay. However, a fairly large fraction of the neutron decays in $^{208}$Pb
would produce at least one neutron emission. These states would produce
$^{207-j}$Pb plus j neutrons, with j being at least one.

\begin{figure}
\includegraphics[angle=-90,width=60ex]{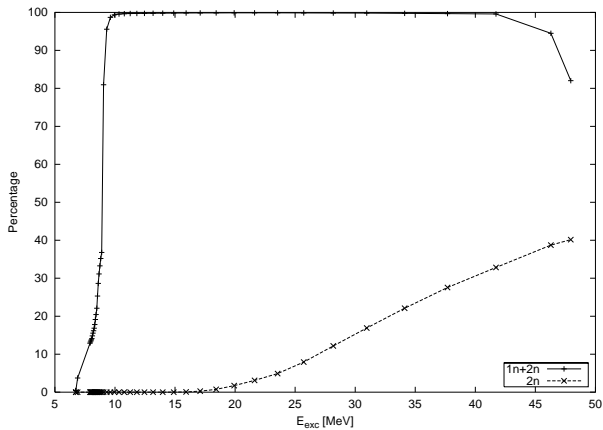}
\caption{\label{fig:pb207} The relative probability for one- and
two-neutron emission as a function of of excitation energy in $^{207}$Pb.}
\end{figure}

The branchings into the particle and $\gamma$-ray emission channels at a given
excitation energy of the daughter nucleus $^{207}$Pb have been calculated.
The relevant transmission coefficients were determined with the same inputs
used in the NON-SMOKER statistical model code \cite{rath00, rath01}, which
is often used for astrophysical calculations, and has been found to be 
accurate over a wide mass range \cite{kola92, rath97}. In addition to 
single-particle emission, two-particle emission can be calculated, 
specifically for the case at hand, the two-neutron emission. For two-particle 
emission each
transition was followed from a level with given spin and parity in
$^{207}$Pb to a definite level with given spin and parity in $^{206}$Pb,
then probabilities for subsequent neutron emission were determined by
summing over neutron emission to all possible final states in $^{205}$Pb. Up
to 20 low-lying experimentally known levels were used in each nucleus 
involved, and a theoretical level density \cite{rath97} was employed above the 
last known state. The relative probability of one- and two-neutron
emission as a function of excitation energy in $^{207}$Pb is shown in Fig.\
\ref{fig:pb207}. Spins from $\frac{1}{2}$ to $\frac{29}{2}$ and both
parities were considered in $^{207}$Pb. The transitions to the states of
different spin were weighted only with $2J+1$. In order to obtain the
correct decay probability, the calculated probabilities have to be folded
with the function that describes how the $^{207}$Pb levels of a given
excitation energy are populated in the primary decay event, as described 
above. The basic
features of the present results will remain although the actual distribution
of strength would be expected to be distorted somewhat due to the folding
with the population derived from $\nu \bar{\nu} \nu$ decay. 

As can be seen from Fig.\ \ref{fig:pb207}, the neutron channel dominates the
possible decays above the neutron emission threshold. Only up to about 9 MeV
is the decay governed by photon emission. Although the proton emission
threshold is at 7.5 MeV, there is no significant proton emission up to 40
MeV in excitation due to the high Coulomb barrier; it is still only 20\% of
all decays at 48 MeV of excitation energy. Although the two-neutron emission
channel opens at 14.83 MeV, it remains insignificant up to 30 MeV in
excitation energy reaching 40\% of all transitions at 48 MeV. It should be
noted that the solid curve in Fig. \ \ref{fig:pb207} contains the sum of
one- and two-neutron emission. Therefore, emissions of single and double
neutrons contribute equally at the highest calculated energy, each
comprising 40\% of all decay possibilities. Thus the population of
excitation energies close to the neutron separation energy can be probed by
measuring $\gamma$-rays, whereas higher excitation energies are accessible
by two-neutron decay.

The lowest lying 82 neutrons in $^{208}$Pb would be sufficiently strongly
bound that their decay would be very likely to populate states at least
above the one-neutron emission threshold. Some of the remaining 44 neutrons
would also be likely to produce daughter states above that threshold. Their
ordering is h$_{9/2}$, f$_{7/2}$, f$_{5/2}$, p$_{3/2}$, i$_{13/2}$, and
p$_{1/2}$, and their approximate binding energies are 12, 11, 8, 8, 8, and 7
MeV respectively \cite{beref}. It seems reasonable to assume that decay of
the neutrons in the last three orbitals would populate states below the
one-neutron emission threshold. Thus 106 of the 126 neutrons in $^{208}$Pb
would be expected to populate states above the one-neutron emission
threshold in $^{207}$Pb. An additional concern, though, is that the decays
of neutrons in the most tightly bound states, \textit{i.e.}, those
dominated by the 1s$_{1/2}$, 1p$_{3/2}$, 1p$_{1/2}$, 1d$_{5/2}$, 2s$_{1/2}$,
and 1d$_{3/2}$ orbitals, might
result in states in $^{207}$Pb that would decay by proton or
$\alpha$-particle emission. Their binding energies are 40, 35, 35, 32, 31,
and 29 MeV respectively \cite{beref}. However, our emission probability
calculations described above suggest that only the states having 1s$_{1/2}$
neutrons would be expected to exhibit such decays; we have subtracted the
two neutrons in that orbital from consideration. Thus 104 neutrons in
$^{208}$Pb can decay to produce detectable signatures.

The upshot is that this nucleon decay mode would have a high probability, 
roughly 104/126 (so that $\epsilon_n$ (see eq. 1) is 104/126=0.825), for 
producing at least one neutron in coincidence with a fairly high-energy 
$\gamma$-ray, but emission of several neutrons in coincidence with a 
$\gamma$-ray would also have a relatively high probability. We have assumed 
50\% for the sum of the probabilities of events in which at least one neutron 
and a detectable ($>$ 3 MeV) $\gamma$-ray are emitted. We note that states 
above 4 MeV of excitation in $^{207}$Pb have a high probability, at least 50\% 
\cite{nucdatasheets}, of producing a $\gamma$-ray of at least 3 MeV in energy. 

We assume 1 kT of lead perchlorate admixed with 20\% water (0.41 kT of lead); 
that has 5.9x10$^{29}$ $^{208}$Pb nuclei, or 6.1x10$^{31}$ neutrons in 
$^{208}$Pb that would be expected to populate states above the one-neutron 
emission threshold of $^{207}$Pb and decay subsequently by neutron emission.
If the fraction of the 104 neutrons assumed that decay by single or multiple 
neutron emission is 100\%, the probability for detection of at least one 
neutron is 50\% (it will be considerably higher for multiple neutron emission), 
the probability for producing a detectable $\gamma$-ray is 50\% and its
detection efficiency is 50\%, then the probability that a neutron decay event 
from one of the 104 neutrons assumed to be detectable will be observed is 
12.5\%. If the lifetime for this decay branch is 10$^{30}$ years then, under
these circumstances, 7-8 events per year would be observed. Note that we have
assumed that only the $^{208}$Pb would contribute; it is likely that the
other lead isotopes would produce similarly detectable decays, producing up
to a factor of 2 enhancement. Furthermore, the assumed efficiencies are
conservative. Thus this mode could have its lifetime extended by searching
for these decays in lead, provided the background events could be managed.

The most obvious background signal that has several neutrons that wouldn't
be vetoed by a cosmic ray shield would involve production of those neutrons
by an energetic cosmic ray in the surrounding rock. One might then get a
$\gamma$-ray if one of the neutrons could inelastically excite a nucleus.
However, this background could be eliminated easily, as the LPC detector
could be surrounded by moderator, so that none of the neutrons getting into
it would have enough energy to inelastically excite a nucleus. 

A more serious background would result from neutrinos produced in the
Earth's upper atmosphere inelastically exciting $^{208}$Pb via the
neutral-current interaction to levels from which it could emit one or more
neutrons, going to nuclei that might themselves emit more neutrons. This
background would be impossible to reject on an event-by-event basis.
However, its yield could be estimated as a function of the energy of the
incident neutrino by measuring the number of charged-current
interactions as a function of energy (as determined by the energy of the
recoiling lepton) that would produce similar numbers of neutrons in
coincidence with the $\gamma$-ray. One could then infer the yield from the
neutral-current interactions on the basis of the relative magnitudes of the
two types of cross sections and the energy
distribution of the charged-current interactions.

\subsection{The n $\rightarrow \ \nu \ \bar{\nu} \ \nu$ decay mode in
$^{35}$Cl.}

The abundance of $^{35}$Cl is 75\% of natural Cl, and there are 2 Cl atoms
per Pb atom in LPC. Thus it is also useful to see if nucleon decay in
$^{35}$Cl might produce a definitive signature.  Neutron decay in $^{35}$Cl
would sometimes produce $^{34}$Cl in a highly excited state. However, the
one-proton emission threshold for $^{34}$Cl is at 5.14 MeV, far below the
one-neutron emission threshold at 11.51 MeV. Indeed, there are apparently no
bound levels in $^{34}$Cl even close to 11.51 MeV. Thus the result of
neutron decay to $^{34}$Cl will be $\gamma$-ray decays to the ground state
of $^{34}$Cl. Note, though, that $^{34}$Cl has an isomeric state at 0.146
MeV; so roughly half of the energetic $\gamma$-rays would go to that state.

\begin{figure}
\includegraphics[angle=-90,width=60ex]{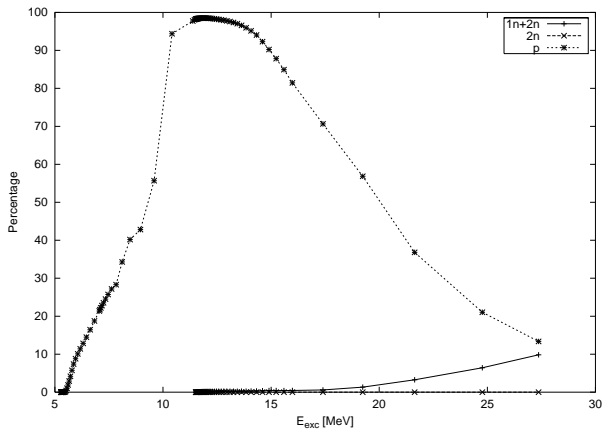}
\caption{\label{fig:cl35} The relative probability for one- and
two-neutron emission and one-proton emission as a function of excitation
energy in $^{34}$Cl.}
\end{figure}

The relative branchings into different decay channels of $^{34}$Cl were
calculated using the same methods as described above for $^{207}$Pb. The
results are shown in Fig.\ \ref{fig:cl35}. The situation for $^{34}$Cl is
different than for $^{207}$Pb, however, because of the lower Coulomb barrier
and a proton separation energy that is lower than the neutron
separation energy. Neutron emission is relatively unimportant at all
calculated excitation energies. The excited nucleus $^{34}$Cl will de-excite
via $\gamma$-transitions for excitation energies up to about 9.5 MeV, where
proton emission accounts for 50\% of all emissions. Proton emission quickly
rises to nearly 100\% by 11.9 MeV and the falls off at high energies due to
increased $\alpha$ emission. Two-neutron emission is completely negligible
at all calculated energies. As with $^{207}$Pb, a quantitative description
of the emission requires the knowledge of the population of the excited
states by the decay, but would not be expected to differ qualitatively from
the results shown here.

The ground state of $^{34}$Cl, however, $\beta$-decays with a half-life of
1.53 s. Thus a coincidence between the energetic $\gamma$-rays (assumed
detection efficiency = 50\%) and the $\beta^+$-decay would identify a
candidate event for neutron decay in $^{35}$Cl. The end point energy of the
$\beta^+$ is 4.47 MeV, so most of the $\beta^+$s will be produced with
sufficient energy to be detected from their Cherenkov radiation (assumed
detection efficiency = 70\%). The isomeric state has a much longer half-life
(32 m), so we have assumed that would not produce a useful coincidence for
determining nucleon decay (reducing the useful event yield by half). None
the less, nucleon decay in $^{35}$Cl could be identified by observing
$\beta$ and $\gamma$s in coincidence, with enough events detected to confirm
the half-life of $^{34}$Cl. The relatively long half-life of $^{34}$Cl would 
demand that the materials used in OMNIS LPC modules be as pure as possible to
minimize accidental coincidences. 

Again assuming 1 kT of lead perchlorate admixed with 20\% water, there will
be 1.8x10$^{30}$ $^{35}$Cl nuclei. Of the 18 neutrons in $^{35}$Cl those in the 
1s$_{1/2}$, 1p$_{3/2}$, and 1p$_{1/2}$ orbits (binding energies = 36, 27, and 
23 MeV respectively \cite{beref}), would probably decay to sufficiently 
highly-excited states in $^{34}$Cl that they would emit a proton,
neutron, or $\alpha$-particle, and not end up in $^{34}$Cl. The remaining
three orbitals, d$_{5/2}$, s$_{1/2}$, and d$_{3/2}$ have binding energies of
roughly 17, 14, and 10 MeV \cite{beref}. The one-neutron separation energy in 
$^{35}$Cl is 12.64 MeV; removal of a valence neutron, presumably from 
a d$_{3/2}$ dominated state, would require that
much energy, and would tend to populate the ground state of $^{34}$Cl. Thus,
presumably, decay of neutrons in $^{35}$Cl states having strong 1d$_{5/2}$ or 
2s$_{1/2}$ neutron configurations would tend to populate states in $^{34}$Cl 
of sufficient excitation energy to emit a detectable, $>$ 3 MeV,
$\gamma$-ray. Thus, about 8
neutrons per $^{35}$Cl nucleus, or 1.4x10$^{31}$ neutrons in the $^{35}$Cl,
(the product N$_n$ $\epsilon_n$ in eq. 1) would have appropriate binding
energy to decay to states in $^{34}$Cl that could produce a $\gamma$-ray above 
3 MeV in energy, so would have a high probability of being detected. If the
detection efficiency for the $\beta$ is assumed to be 70\%, and the
lifetime for this decay process is 10$^{30}$ years, one would expect to observe
3-4 events per year from neutron decay in $^{35}$Cl. Although this decay mode
would not produce as strong a limit on nucleon decay as would the lead, its
signature would be considerably more definitive than that from the lead,
because its backgrounds are so much more readily identifiable.

The primary background for decay of a neutron in $^{35}$Cl comes from 
neutrinos produced in the Earth's atmosphere, but such events would not be
able to simulate nucleon decay events. Although once the neutron was knocked
out of the $^{35}$Cl, the resulting $\gamma-\beta$ coincidence would be the
same as for the neutron decay event, the atmospheric neutrino would emit a
neutron, which would be detected to give its
characteristic 8.6 MeV $\gamma$-ray. This is a considerably greater energy
than could be produced by any of the $\gamma$-rays resulting from the
de-excitations in $^{34}$Cl; thus that neutron could be used to veto this
type of background event.

\subsection{Monte-Carlo simulations of detectors to determine detection
efficiencies}

The limit that can be achieved from any process such as the decays described
above clearly depends on the detection efficiency. Thus we have run
simulations of the events from decays of $^{35}$Cl based on
the GEANT~\cite{geant} detector simulation software. As planned, the LPC
detector will consist of multiple independent cylindrical modules having
radii of 2 m and heights, which are adjustable in the simulations, ranging
from 1 to 2 m. The GEANT simulations were based on a single module. The
interior of the cylinder was assumed to be viewed by photomultiplier tubes
(PMTs) at  both ends. Current design plans are to use an LPC solution of
concentration, also adjustable in the simulations, of 75\% to 80\% LPC by
weight.

From the decay of a neutron in $^{35}$Cl, GEANT generated both the 
$\gamma$-rays from the de-excitation of the $^{34}$Cl nucleus and the 
resultant $\beta^+$ from the decay of the $^{34}$Cl. The initial starting
state for $^{34}$Cl was randomly chosen from the energy levels that were
below the proton separation energy. $\gamma$-rays were then generated to
simulate the de-excitation of the $^{34}$Cl nucleus from the selected
starting state down to either the ground state or the isomer. The number and
energies of the generated $\gamma$-rays were 
based on data on the levels and $\gamma$-rays for the $^{34}$Cl nucleus.
Events that decayed to the isomeric state were rejected. The fraction of the
events that de-excite down to the ground state was found to be $57.6 \pm
0.2$\%.

The $^{34}$Cl decay time is chosen by randomly selecting whether a decay
will occur during the 1 ms time unit, and was repeated until a decay
occurred. The chance for a decay within a time unit was based on the
half-life of $^{34}$Cl. If the decay time took longer than 40 s, the total
accumulated time was set to zero and the decay test continued.

The kinematics for the $\beta^+$ were generated after the $\gamma$-rays 
were generated and tracked. The initial energy of the $\beta^+$ was selected
from a lookup table and the decay time was added to the initial time of
flight. A Gaussian profile was used for the initial energy distribution of
the positron. The parameters for the Gaussian were derived from a fit to the
$\beta^+$ energy spectrum calculated using the RADLST program~\cite{radlst}.

Upon entering a PMT, a photon's energy and time of flight were stored. As
$\beta^+$s and $\beta^-$s have a large RMS multiple scattering in LPC, any
patterns in the Cherenkov radiation are destroyed. However, an event can be
identified by the number of PMTs that have fired in a localized area within
a specified time window. The requirement imposed to identify $\gamma$-rays
and $\beta^+$s required three PMTs to fire in a localized cluster within a
20 ns window.

To estimate the efficiency for detecting neutron decay in the $^{35}$Cl in
LPC, 10,000 events were generated in a 100 cm deep tank with an 80\%
solution of LPC by weight. The attenuation length of the LPC was taken to be
4.2 m \cite{elliott00}. The mean efficiency for detecting the $\gamma$-rays
from the de-excitation of the $^{34}$Cl nucleus over the volume of the
detector was found to be $57.0 \pm 0.8$\%. For detecting the $\beta^+$s, the
mean efficiency was found to be $75.2 \pm 0.9$\%, making the mean efficiency
for detecting a coincidence between the $\gamma$s and the $\beta$s to be
$42.6 \pm 
0.8$\%. For a 200 cm deep tank with an 80\% LPC solution, the mean detection
efficiency for $\gamma$s, $\beta^+$s, and coincidences was found to be $42.6
\pm 0.7$\%, $60.6 \pm 0.9$\%, and $25.6 \pm 0.5$\% respectively. Thus the
efficiencies assumed above are consistent with those determined from the
Monte-Carlo simulations. For the 200 cm deep tank, the neutron detection
efficiency was found to be $87.6\pm 0.9$\%.

The position of a nucleon decay event can be deduced from the time
difference $\Delta$t between hits at the two ends of the tank. $\Delta$t can
be defined in a variety of ways, \textit{e.g.} by taking it to be the time
difference between the peaks of the signals, either from the $\gamma$s or
the $\beta^+$s, from the two sides of the module, or by attempting to
average early hits to utilize the leading edges of the signals on the two
sides to do the timing. The first approach can be confused somewhat by the
arrival of photons scattered from the 
opposite side, whereas the latter can be complicated by limited statistics.
Either approach seems to allow localization of events to about $\pm 10$ cm.
Spatial localization is important, as it allows use of a position 
dependent
neutron detection efficiency, so can allow greater emphasis on events that
occur in the center of the detector, which will have a greater efficiency
for detecting the veto neutron, 
than those at the edges, which will have a lesser efficiency.

\section{Conclusions}

Given that the suggested searches for nucleon decay would be conducted in a
supernova neutrino detector, these searches would have a long time, probably
more than 20 years, to run. Thus the suggested experiments have the
potential to improve greatly the existing limits on the n $\rightarrow \ \nu
\ + \bar{\nu} \ + \nu$ decay mode. 

\begin{acknowledgments}

This work began at an Aspen Institute summer workshop. It has continued
under the support of US Department of Energy grant DE-FG03-88ER40397, 
NSF grant PHY-0099476, and the Swiss NSF (grant 2000-061031.02). TR
acknowledges support by a PROFIL professorship by the Swiss NSF
(grant 2024-067428.02). 

\end{acknowledgments}


\begin{thebibliography}{}

\bibitem{pati73}
J.C. Pati and A. Salam, Phys. Rev. D \textbf{8}, 1240 (1973); Phys. 
Rev. Lett. \textbf{31}, 661 (1973); Phys. Rev. D \textbf{10}, 275 (1974).
\bibitem{georgi74}
H. Georgi and S.L. Glashow, Phys. Rev. Lett. \textbf{32}, 438 (1974).
\bibitem{babu00}
K.S. Babu, J.C. Pati, and F. Wilczek, Nucl. Phys. \textbf{B566}, 33 (2000).
\bibitem{babu01}
K.S. Babu and R.N. Mohapatra, Phys. Lett. \textbf{B518}, 269 (2001).
\bibitem{applequist01}
T. Appelquist, B.A. Dobrescu, E. Ponton, and H.-U. Yee, Phys. Rev. Letters
\textbf{87}, 181802 (2001).
\bibitem{pdg}
H. Hagiwara \textit{et al.}, Particle Data Group, Phys. Rev. D \textbf{66},
010001 (2002); http://pdg.lbl.gov 
\bibitem{totsuka86}
Y. Totsuka, in \textit{Proceedings of the 7th Workshop on Grand 
Unification/ICOBAN, 1986}, edited by J. Arafune (World Scientific,
Singapore, 1986), p. 118
\bibitem{ejiri93}
H. Ejiri, Phys. Rev. C \textbf{48}, 1442 (1993).
\bibitem{suzuki93}
Y. Suzuki \textit{et al.}, Phys. Lett. \textbf {B311}, 357 (1993).
\bibitem{kamyshkov02}
Y. Kamyshkov and E. Kolbe, Phys. Rev. D \textbf{67}, 076007 (2003).
\bibitem{pati83}
J. Pati, A. Salam, and U. Sarker, Phys. Letters \textbf{B133}, 330 (1983).
\bibitem{mohapatra02}
R.N. Mohapatra and A. Perez-Lorenzana, Phys. Rev. D \textbf{67}, 075015 
(2003).
\bibitem{berger91}
C. Berger \textit{et al.}, Phys. Letters \textbf{B269}, 227 (1991).
\bibitem{boyd02}
R.N. Boyd, A. StJ. Murphy, and R.L. Talaga, Nucl. Phys. \textbf{A718}, 222c
(2003).
\bibitem{elliott00}
S.R. Elliott, Phys. Rev. C \textbf{62}, 065802 (2000).
\bibitem{nucdatasheets}
Evaluated Nuclear Structure Data File, http://www.nndc.bnl.gov/nndc/ensdf
\bibitem{rath00}
T. Rauscher and F.-K. Thielemann, At. Data Nucl. Data Tables \textbf{75}, 1
(2000)
\bibitem{rath01}
T. Rauscher amd F.-K. Thielemann, At. Data Nucl. Data Tables \textbf{79}, 47
(2001)
\bibitem{kola92}
E. Kolbe, K. Langanke, S. Krewald, and F.-K. Thielemann, Nucl. Phys.
\textbf{A540}, 599 (1992)
\bibitem{rath97}
T. Rauscher, F.-K. Thielemann, and K.-L. Kratz, Phys. Rev. \textbf{C56},
1613 (1997)
\bibitem{beref}
A. Bohr and B.R. Mottelson, \textit{Nuclear Structure}, (World Scientific,
Singapore,1998), p. 239
\bibitem{geant}
GEANT 3.21, CERN Program Library, Long Writeup W5013 (1993).
\bibitem{radlst}
T.W. Burrows, computer code RADLST, Brookhaven National Laboratory 
Report No. BNL-NCS-52142, 1988 (unpublished).

\end{thebibliography}
\end{document}